\documentclass[doublecol]{epl2} 
\usepackage{amsmath}
\usepackage{cite}

\usepackage[normalem]{ulem}

\title{Effective medium theory of random regular networks}

\author{O. K. Damavandi\inst{1} \and M. L. Manning\inst{1} \and J. M. Schwarz\inst{1,2}}
\shortauthor{O. K. Damavandi \etal}

\institute{                    
  \inst{1} Department of Physics and BioInspired Institute, Syracuse University, Syracuse, New York 13244, USA\\
  \inst{2} Indian Creek Farm, Ithaca, New York 14850, USA
}

\abstract{
Disordered spring networks can exhibit rigidity transitions, due to either the removal of materials in over-constrained networks or the application of strain in under-constrained ones. While an effective medium theory (EMT) exists for the former, there is none for the latter. We, therefore, formulate an EMT for random regular networks, under-constrained spring networks with purely geometrical disorder, to predict their stiffness via the distribution of tensions. We find a linear dependence of stiffness on strain in the rigid phase and a nontrivial dependence on both the mean and standard deviation of the tension distribution. While EMT does not yield highly accurate predictions of shear modulus due to spatial heterogeneities, the noninvasiveness of this EMT makes it an ideal starting point for experimentalists quantifying the mechanics of such networks. 
}

\begin{document}

\maketitle

\section{Introduction}

Elasticity theory---both its linear and nonlinear forms---is a continuum approach to understand the mechanics of materials~\cite{Landau,Ogden84}. While elasticity theory can provide accurate predictions regarding the deformability of many materials ranging from carbon nanotubes to biological tissues~\cite{Tibbetts84,Engstrom19}, it is not readily equipped to characterize disordered materials and how they potentially lose their elasticity when undergoing a rigidity transition.  And yet, the phenomenon of a rigidity transition is rather common in systems where the density of the disordered material is modified, such as tuning the packing fraction in granular packings~\cite{liu_jamming_2010,VanHecke2009}, or the removal, or fracture, of material in disordered spring networks~\cite{Feng1984,Berthier19}. In the latter example, the spring network goes from over-constrained, where the number of degrees of freedom are fewer than the number of springs imposing independent constraints, to under-constrained and, hence, the loss of rigidity. One of the main theoretical frameworks characterizing density-dependent rigidity transitions is known as rigidity percolation. In rigidity percolation, one looks for the loss of a spanning rigid cluster, or the vanishing of elastic moduli, as springs are randomly removed from the network~\cite{Feng1984,Feng84b,Jacobs95}.

Despite the minimal framework of rigidity percolation, analytical approaches remain mostly limited to spring networks in the absence of loops~\cite{Duxbury99,Barre09}.  In such networks, however, the nature of the rigidity transition, in terms of continuous or discontinuous or a hybrid of the two, depends on some details of the network, making it difficult to "see" how a continuum approach to rigidity percolation would work~\cite{Jacobs95,Moukarzel97,Chubynsky07}. However, one approximate analytical approach is Effective Medium Theory (EMT)~\cite{Feng85,Tang88}. EMT maps a disordered spring network to an ordered spring network using effective spring constants that are obtained by quantifying the response of a nonrandom spring network to the replacement of one spring with a different spring constant than the surrounding network, or medium. This medium does not fluctuate by construction. In this sense, EMT is a mean field theory for rigidity percolation and indeed captures qualitative aspects of the rigidity transition as well as even exponents associated with the transition~\cite{Feng85,Tang88}. Given the approximate nature of the approach, the obtained exponents do not necessarily agree with numerical simulations of the rigidity transition~\cite{He1985}.  More recent work has extended EMT to include angular springs~\cite{Das2012,Mao2013} and anisotropic networks\cite{Zhang14}. There also exists a nonlinear extension as well~\cite{Sheinman2012}.

Interestingly, rigidity transitions occur in spring networks even when the amount of material stays fixed~\cite{Sharma2016,Merkel2019}. These spring networks typically start out as floppy, or not rigid, in the absence of any perturbation but then can become rigid as, for example, an applied shear is large enough such that some backbone of springs become taut, thereby exhibiting what is know as a self-stress. This phenomenon can also occur in some versions of vertex models, which are multi-body generalizations of two-body spring networks, as a function of increasing internal strain by decreasing equilibrium spring lengths~\cite{Bi2015,Merkel2019}. Researchers have also studied the rigidity transition in vertex models when the internal strain and the connectivity is varied~\cite{Yan2019,Petridou21}. These density-independent rigidity transitions have a range of applicability, particularly in biology, where disordered two-body and three-body spring networks are realizations of the biopolymer networks, such as the extracellular matrix (ECM), and vertex models are realizations of confluent tissues. 

Given the range of applicability for the density-independent rigidity transitions in under-constrained networks, a natural, pressing question emerges regarding what types of analytical approaches can be used to predict the emergent mechanics in such a transition.  Since EMT exists for emergent mechanics for networks with increasing connectivity (or material), can one construct an EMT for disordered spring networks undergoing a rigidity transition via a change in internal or external strain and with fixed connectivity?  Here, we answer in the affirmative and present the construction for random, regular spring networks---ones in which each vertex has the same local connectivity but a distribution of equilibrium spring lengths which, in turns, leads to a distribution of tensions. In other words, the source of the disorder, here, is purely geometrical and can be tuned in ways to vary the hetereogeneity of the network. We test our analytical construction against numerical simulations. Additionally, we ask: Can we tell if a network will be stiffer than another network by just looking at the distribution of tensions? In other words, does the tension distribution contain enough information to qualitatively predict the elastic behavior? We will begin to answer this question by comparing our EMT with numerical simulations. Finally, the development of this EMT enables us to map disordered networks to ordered ones without prior knowledge of the source of the disorder but by just looking at the distribution of tensions, which would be useful in experiments in which data exists only in the form of network snapshots.

\section{Numerical model for spring networks}
\begin{figure}
\centering
    \includegraphics[width=\linewidth]{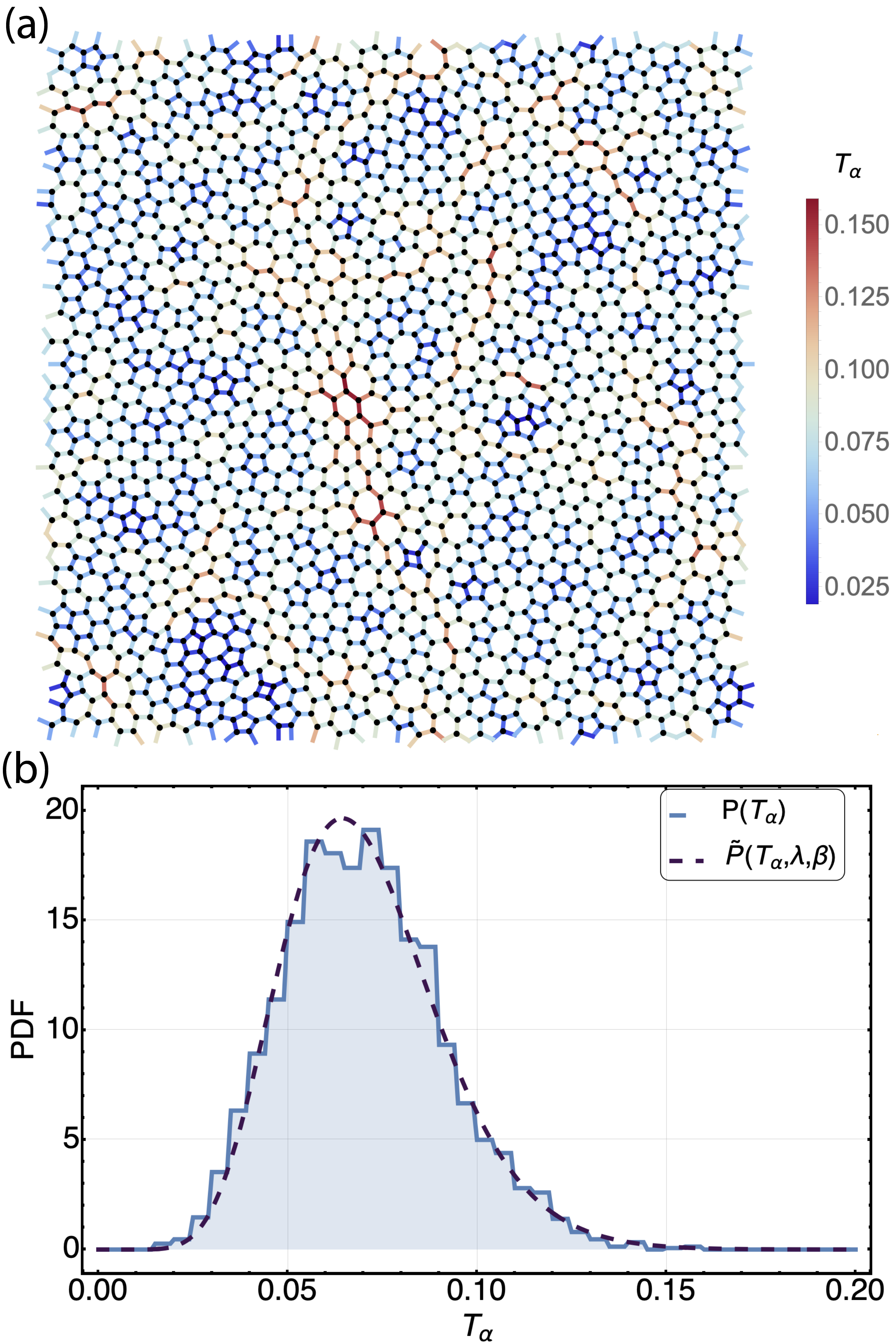}
\caption{(a) A rigid random regular network with $N=2000$ vertices and $L_0 = 0.550$ in a periodic box with $L_{box} = \sqrt{1000}$ (for this network $L_0^*\approx0.628$). The colors show the tensions on the edges. (b) Probability distribution functions (PDF) for tension distribution $P(T_\alpha)$ for the same network and the fitted Gamma distribution, $\widetilde{P}(T_\alpha,\lambda,\beta)$. For this network, $\bar{T}=0.071$ and $\sigma_T = 0.021$.}
\label{fig:random_regular_net}
\end{figure}

Let us first detail how our random, regular spring networks are generated. The energy of the spring network is given by

\begin{equation}
    E = \frac{k}{2} \sum_{\alpha=1}^{3N/2} (L_\alpha - L_0)^2,
    \label{eq:energy}
\end{equation}
where $\alpha$ labels the edges and $L$ denotes the length of a spring. Without loss of generality we have assumed all springs have the same rest length $L_0$~\cite{Merkel2019}. Numerically, we generate a random regular spring network with $N$ vertices by placing $N/2$ cell centers in a periodic box and identifying the unique Voronoi tessellation associated with that set of points~\cite{Sussman2017}. In two dimensions (2D), all vertices of this network will have coordination number $z=3$ (Fig.~\ref{fig:random_regular_net}a). If the initial set of cell centers is generated randomly from a uniform distribution, this process will generate a disordered network with quenched geometric disorder as opposed to randomness in coordination number $z$ as is the case in random diluted networks~\cite{Jacobs95,Feng85}. Unlike random diluted networks, the randomness of a random regular network is in the number of edges that each cell has (see Fig.~\ref{fig:random_regular_net}a) and does not show up explicitly in the energy.

The resulting network (Fig.~\ref{fig:random_regular_net}a) is highly under-coordinated and, thus, using constraint-counting arguments, one would expect it to be floppy. However, as one decreases $L_0$, the system undergoes a rigidity transition at a particular $L_0^*$ as typically indicated by the emergence of a non-zero shear modulus due to the onset of geometric frustration~\cite{Merkel2019}. In other words, $L_0 < L_0^*$, springs have nonzero tensions because they cannot reach their equilibrium spring length, the shear modulus $G>0$ and thus the system is rigid while for $L_0 > L_0^*$, all springs are at their rest length, $G=0$,  and the system is floppy~\cite{Damavandi2021,Merkel2019}. We can also see that in the rigid regime, the geometric disorder is reflected in the distribution $P(T_\alpha)$ of edge tensions $T_\alpha = k(L_\alpha - L_0)$. See Fig. 1b. It has also been shown that such networks exhibit a rigidity transition at a system dependent $L_0^*$. Such system dependency is common in phase transitions and is typically due to finite size effects. See Supplementary material (SM).

We can encode different degrees of spatial heterogeneity in the network by applying a self-propelled Voronoi (SPV) dynamics~\cite{Sussman2017} for different number of time steps $T_{spv}$ after Voronoi tessellation but before energy minimization (more in SM). With this method, we generate two networks: Network 1 with $T_{spv}=100$ and network 2 with $T_{spv}=5$. As a result, network 1 is more spatially uniform as reflected in $P(T_\alpha)$ (Fig.~\ref{fig:random_regular_net}b and \ref{fig:EMT}a). We also observe that $P(T_\alpha)$ is well-fitted by a Gamma distribution, i.e.\ if $\bar{T}$ and $\sigma_T$ are the average and standard deviation of edge tensions respectively, we find $P(T_\alpha)\approx \widetilde{P}(T_\alpha,\lambda,\beta)= e^{-T_\alpha/\beta}T_\alpha ^{\lambda-1}\beta^{-\lambda}/\Gamma(\lambda)$ where $\lambda= (\bar{T}/\sigma_T)^2$ and $\beta= \sigma_T^2/\bar{T}$ (Fig.~\ref{fig:EMT}a). 

Finally, $P(T_\alpha)$ implicitly depends on how much global strain is applied to the springs. We can define $u_0 = L_0^* - L_0$ to be the global isotropic expansion that is applied to the disordered network. Merkel {\it et al.}~\cite{Merkel2019} showed that close to $L_0^*$, the average edge length $\bar{L}$ linearly scales with $\sigma_T$: $\bar{L} = a_L \sigma_T + L_0^*$ where $a_L$ is a system dependent scaling factor. We can use this relationship to extrapolate $L_0^*$ numerically by generating multiple rigid networks with the same initial geometry but with different $L_0$'s close to the rigidity transition, fitting a line and extracting the y-intercept. With this, we can find $u_0$ for each network.

Our goal is to map random regular networks to an ordered network with the same topology (honeycomb lattice in $2D$) by developing an EMT using $P(T_\alpha)$ as the only set of information about the network. The question we address is whether the tension distribution encompasses enough information about the geometric disorder for the EMT to accurately predict the shear modulus of a given disordered network.

\section{Theoretical results}
\begin{figure*}
\begin{center}
    \includegraphics[width=0.64\linewidth]{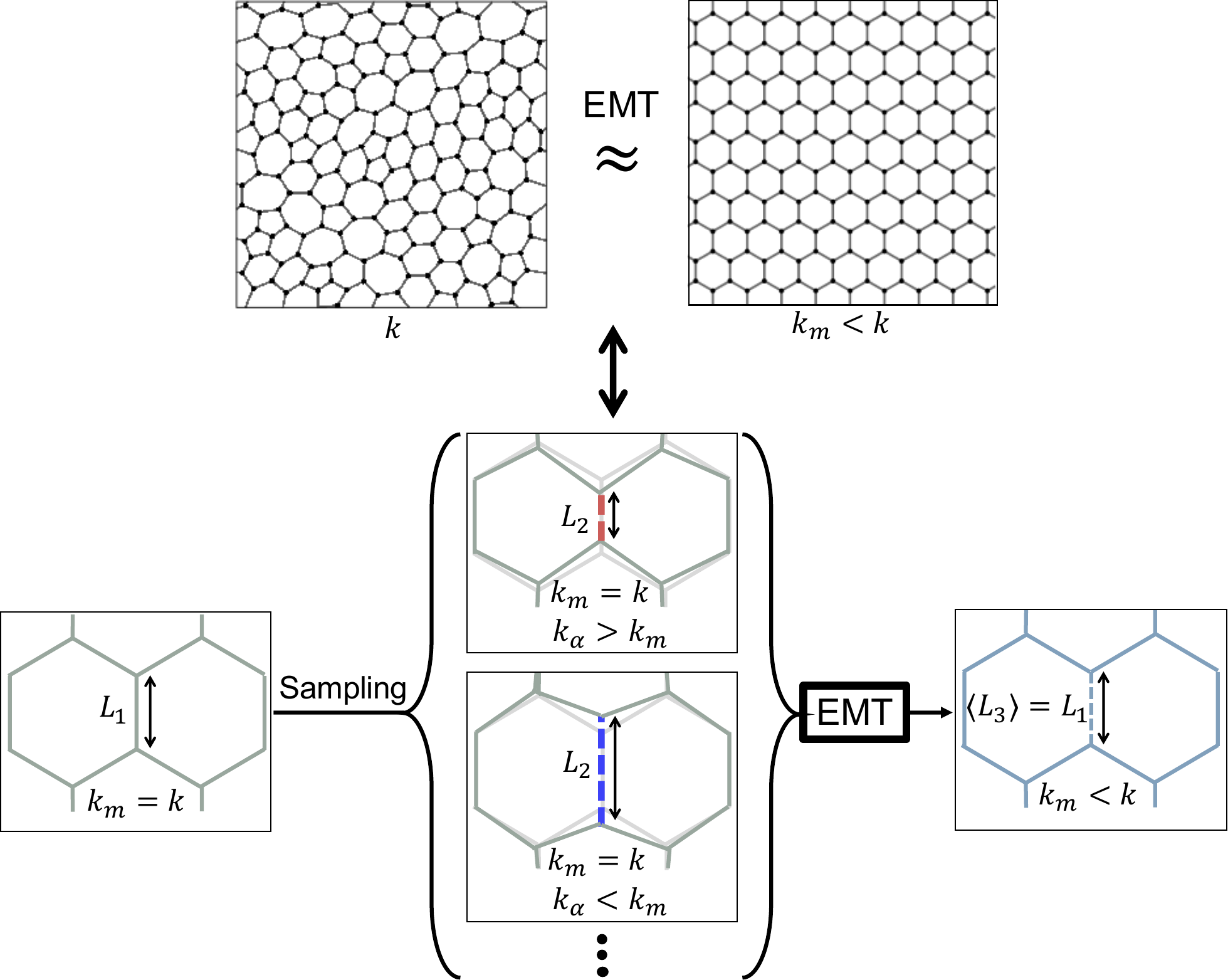}
\caption{A schematic of the EMT process. After the honeycomb lattice with $k_m = k$ has gone under a uniform expansion $u_0$ and all the edge lengths are $L_1 = L_0^* + u_0$, we replace one edge by an edge with spring constant $k_\alpha$ that causes a deformation in the network such that the edge is displaced by $\delta u$, $L_2 = L_1 + \delta u$. Two possibilities for $L_2$ have been shown, one where $k_\alpha<k_m$ (blue) and one where $k_\alpha>k_m$ (red). The EMT process seeks to find a $k_m$ (typically less than $k$) such that $\langle \delta u \rangle =0$ or the length of the wrong bond after the process, $\langle L_3 \rangle =L_1$.}  
\label{fig:EMT_schematic}
\end{center}
\end{figure*}

Here, we develop an EMT for the disordered network in the rigid regime ($L_0<L_0^*$). If we apply $u_0$ to an ordered (honeycomb) network with spring constant $k_m$, all edges will have tension $T_m = k_m u_0$. But a disordered network will show nonaffine deformations around $u_0$ captured in $P(T_\alpha)$. Now, replace an edge in the honeycomb with an edge with spring constant $k_\alpha (u_0) = T_\alpha/u_0$ where $T_\alpha$ is sampled from $P(T_\alpha)$ (Fig.~\ref{fig:EMT_schematic}). If this wrong edge is held at displacement $u_0$ so that the honeycomb remains undeformed, it will hold tension $T_\alpha$ instead of $T_m$. We want to find $k_m$ that makes the deformation $\delta u$ due to the edge swap vanish on average, $\langle \delta u \rangle =0$
where $\langle \dots \rangle$ is an ensemble average over the sampled $P(T_\alpha)$, such that the wrong edge with tension $T_\alpha$ is supported by the the rest of the ordered network with tension $T_m$. This effective ordered network would then be expected to possess the same elastic properties (e.g.\ shear modulus) as the disordered network.

We can find a self consistency equation for $k_m$ by writing $\delta u = f/\bar{k}$ where $f$ is the force causing deformation $\delta u$, or equivalently, the external force needed to undo $\delta u$: $f = T_m - T_\alpha$. $\bar{k}$ is the spring constant felt by the vertices connected by the edge that is being swapped, and it can be shown~\cite{Tang88} that $\bar{k} = k_m/a^* - k_m + k_\alpha$. Moreover, $a^*$ is given by~\cite{Das2012,Liarte2016}

\begin{equation}
    a^* = \frac{1}{z} \int_{\mathrm{1BZ}}  \frac{\mathrm{d}\mathbf{q}}{\tilde{v}_0}\, \mathrm{Tr}(\mathbf{D}_{g,\mathbf{q}}\cdot \mathbf{D}_{\mathbf{q}}^{-1}),
    \label{eq:astar}
\end{equation}
where $\mathbf{D}_{\mathbf{q}} = (1-u_0/L_0^*)\mathbf{D}_{g,\mathbf{q}}+(u_0/L_0^*) \mathbf{D}_{t,\mathbf{q}}$ is the $4\times 4$ dynamical matrix in Fourier space derived from the energy of small perturbations of the two vertices in each unit cell $\mathrm{d}E=(1/2N_{cell})\sum_\mathbf{q} (\begin{smallmatrix}\mathbf{u}_{-\mathbf{q}},& \mathbf{v}_{-\mathbf{q}}\end{smallmatrix}) \cdot \mathbf{D}_{\mathbf{q}} \cdot \left(\begin{smallmatrix}\mathbf{u}_{\mathbf{q}}\\ \mathbf{v}_{\mathbf{q}}\end{smallmatrix}\right)$, while  $\mathbf{D}_{g,\mathbf{q}}$ is the geometric part of $\mathbf{D}_{\mathbf{q}}$ and $\mathbf{D}_{t,\mathbf{q}}$ is the part of $\mathbf{D}_{\mathbf{q}}$ due to the tension in the network. 
Analogous to a lattice under tension with one vertex per unit cell (discussed in ref.~\cite{Tang88}) we derive

\begin{equation}
    \mathbf{D}_{g,\mathbf{q}} = \sum_{j=1}^3 
    \begin{pmatrix}
        \mathbf{n}_j \otimes \mathbf{n}_j & -e^{-i \mathbf{q}\cdot \mathbf{a}_j} \mathbf{n}_j \otimes \mathbf{n}_j\\
        -e^{i \mathbf{q}\cdot \mathbf{a}_j} \mathbf{n}_j \otimes \mathbf{n}_j & \mathbf{n}_j \otimes \mathbf{n}_j
    \end{pmatrix}
\end{equation}
and 

\begin{equation}
    \mathbf{D}_{t,\mathbf{q}} = \sum_{j=1}^3
    \begin{pmatrix}
        \mathbf{I}_{2\times 2} & -e^{-i \mathbf{q}\cdot \mathbf{a}_j} \mathbf{I}_{2\times 2}\\
        -e^{i \mathbf{q}\cdot \mathbf{a}_j} \mathbf{I}_{2\times 2} & \mathbf{I}_{2\times 2}
    \end{pmatrix}
\end{equation}

\begin{figure}
\centering
    \includegraphics[width=0.5\linewidth]{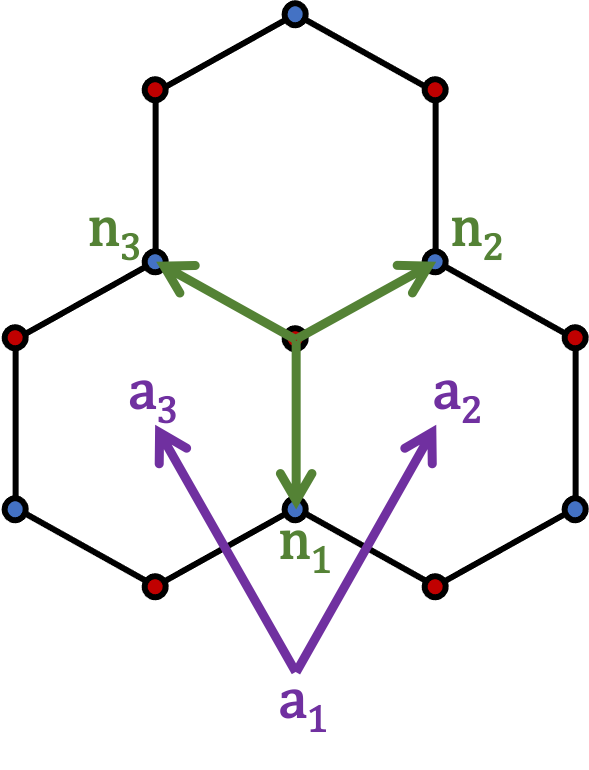}
\caption{A honeycomb lattice showing the vectors $\mathbf{n}_i$ and $\mathbf{a}_i$. }
\label{fig:honeycomb}
\end{figure}

Here, $n_j$ and $a_j$ are lattice vectors defining the honeycomb lattice and its dual respectively: $\mathbf{a}_1=(0,0),\, \mathbf{a}_2=(\sqrt{3}/2,3/2),\, \mathbf{a}_3=(-\sqrt{3}/2,3/2),\, \mathbf{n}_1=(0,-1),\, \mathbf{n}_2=(\sqrt{3}/2,1/2),\, \mathbf{n}_3=(-\sqrt{3}/2,1/2)$ (Fig.~\ref{fig:honeycomb}), and $\mathbf{I}_{2\times 2}$ is the $2\times 2$ identity matrix. Since the honeycomb lattice is under-coordinated, $\mathbf{D}_{g,\mathbf{q}}$ is singular, but under tension the honeycomb is rigid and $\mathbf{D}_{\mathbf{q}}$ is invertible. 

Putting everything together, we derive the EMT self-consistency equation, $\langle \delta u \rangle =0$ averaged over the tensions sampled from $P(T_\alpha)$:

\begin{equation}
    \int \frac{k_m u_0 - T_\alpha}{(k_m/a^* - k_m)u_0 + T_\alpha} P(T_\alpha) = 0.
    \label{eq:self_consistency}
\end{equation}
This equation solves for $k_m(u_0)$ given a global expansion $u_0$ and a tension distribution $P(T_\alpha)$ (itself an implicit function of $u_0$), which can be used to find the effective medium shear modulus approximation of the disordered network

\begin{equation}
    G_{m}(u_0) = \frac{k_m(u_0)}{k} G_{h.c.}(u_0).
    \label{eq:G_m}
\end{equation}
Here, $G_{h.c.}$ is the shear modulus of a honeycomb lattice with spring constant $k$. Note that unlike the previous iterations of EMT, the self consistency equation directly uses the observed tension distribution instead of an underlying spring constant distribution. This allows for more accessible application to experimental data. 
 
\section{Numerical results}
\begin{figure}
\centering
    \includegraphics[width=\linewidth]{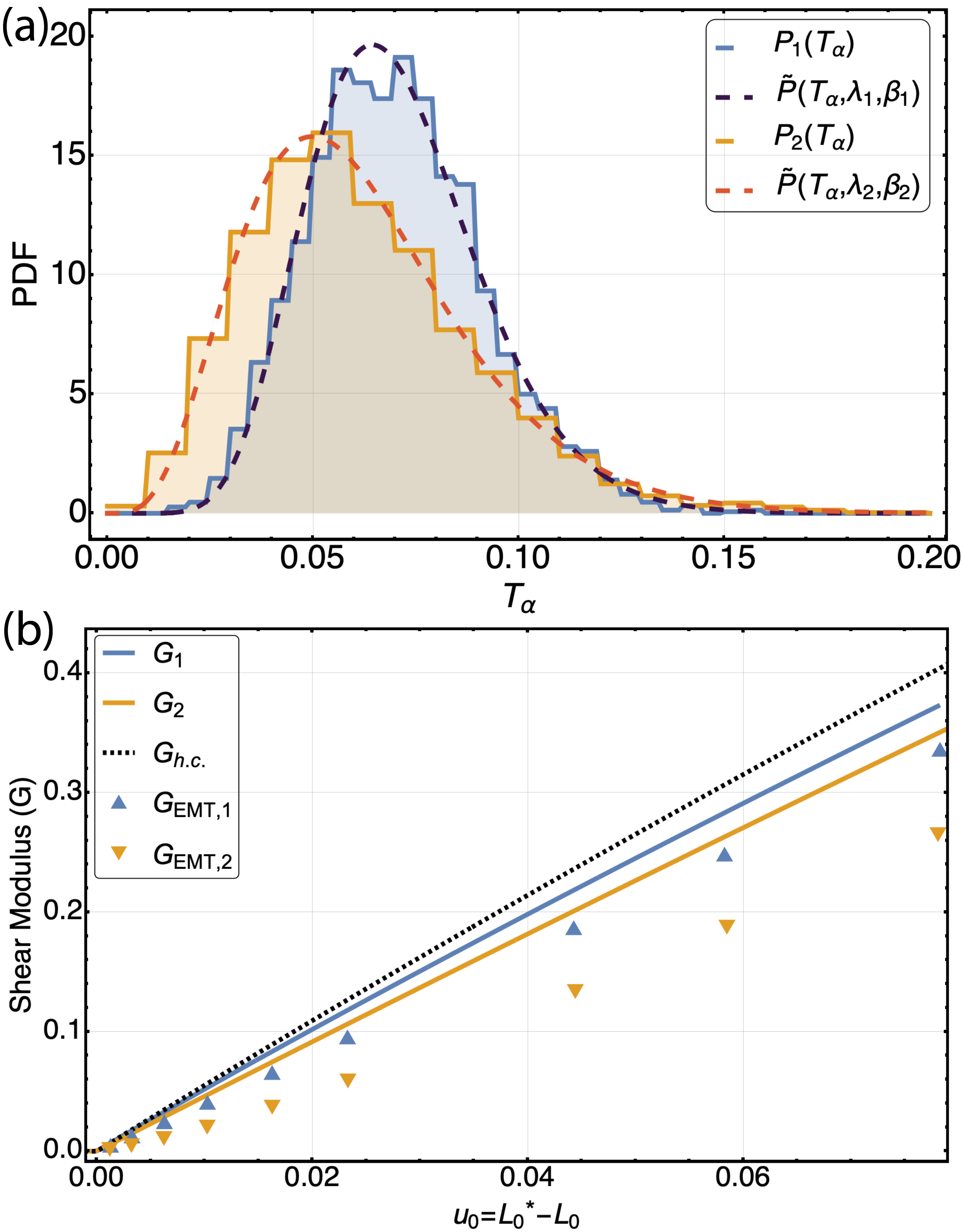}
\caption{(a) Probability distribution functions (PDF) for network 1 ($\bar{T}_1=0.071$ and $\sigma_{T,1} = 0.021$) and network 2 ($\bar{T}_1=0.062$ and $\sigma_{T,2} = 0.028$) both at $u_0=0.078$. (b) Shear modulus $G$ vs.\ $u_0$ for network 1 and 2 and their respective EMT. $G_{h.c.}$ with the same spring constant as the disordered networks has also been plotted for reference. EMT correctly predicts that network 2 must have a lower shear modulus.}
\label{fig:EMT}
\end{figure}

We numerically calculate the shear modulus for a disordered network (network 1) and compare with the EMT prediction (Fig.~\ref{fig:EMT}b) using ~\cite{Merkel2018} 

\begin{align}
    G &= \frac{1}{N_{cell}}\frac{\mathrm{d}^2E}{\mathrm{d}\gamma^2}\nonumber\\
    &=\frac{1}{N_{cell}}\left(\frac{\partial^2E}{\partial \gamma^2}-\sum_l\frac{1}{\lambda_l}\left[\sum_n\frac{\partial^2E}{\partial \gamma \partial x_n}u^{(l)}_n\right]\right),
    \label{eq:shearModulus}
\end{align}
where $\gamma$ is the infinitesimal shear parameter, $\lambda_l$ and $u^{(l)}$ are the eigenvalues and eigenvectors of the Hessian matrix and $x_n$ is the $n$th degree of freedom (position of vertices). We observe that EMT consistently under-estimates the shear modulus of the disordered network but it qualitatively gets the trend correct. The quantitative disagreement should not come as a surprise because EMT is an uncontrolled approximation. Moreover, there are spatial correlations in the disorder (Fig.~\ref{fig:random_regular_net}a) that are not necessarily captured by $P(T_\alpha)$. There are, in fact, network-spanning, chain-like structures that carry high tensions (see SM). Without these structures that hold the network together under strain, the network would be less rigid. Since these spatially organized structures are not explicitly encoded in $P(T_\alpha)$, EMT treats high tension edges as uniformly distributed instead, and thus predicts a lower shear modulus.

We now ask: given the lack of quantitative agreement, can EMT still be useful to compare stiffnesses of different networks with different amounts of geometric heterogeneity merely from their tension distributions? To answer this question, we compare EMT predictions for network 1 (with $T_{spv} = 100$) and network 2 (with $T_{spv} = 5$). We can see that network 2 has a higher degree of spatial heterogeneity than network 1 as captured by a wider $P(T_\alpha)$ (Fig.~\ref{fig:EMT}a). Network 2 also shows a lower shear modulus than network 1 (Fig.~\ref{fig:EMT}b) agreeing with the intuition that more spatial heterogeneity, which leads to larger nonaffine displacements, should lower $G$. Applying EMT to both networks, we see that it also predicts a lower shear modulus for network 2 (Fig.~\ref{fig:EMT}b). However, again the EMT predictions are quantitatively not in great agreement with the numerical results. Notably, EMT for network 2 shows less of an agreement compared to network 1. This is potentially due to the fact that network 2 has a higher spatially correlated disorder compared to network 1. To better quantify spatial correlations, we look at the degree of local nematic ordering in the high tension edges (i.e.\ edges with higher tension $>\bar{T}$) measured by a scalar nematic order parameter $\bar{S}$ averaged over the network (see SM for more details). We expect more chain-like alignments in network 2 and thus a larger $\bar{S}$. Indeed we see $\bar{S}_1 = 0.379 \pm 0.005$ and $\bar{S}_2 = 0.431 \pm 0.018$ (averaged over three networks with the same $T_{spv}$).

\begin{figure}
\centering
    \includegraphics[width=\linewidth]{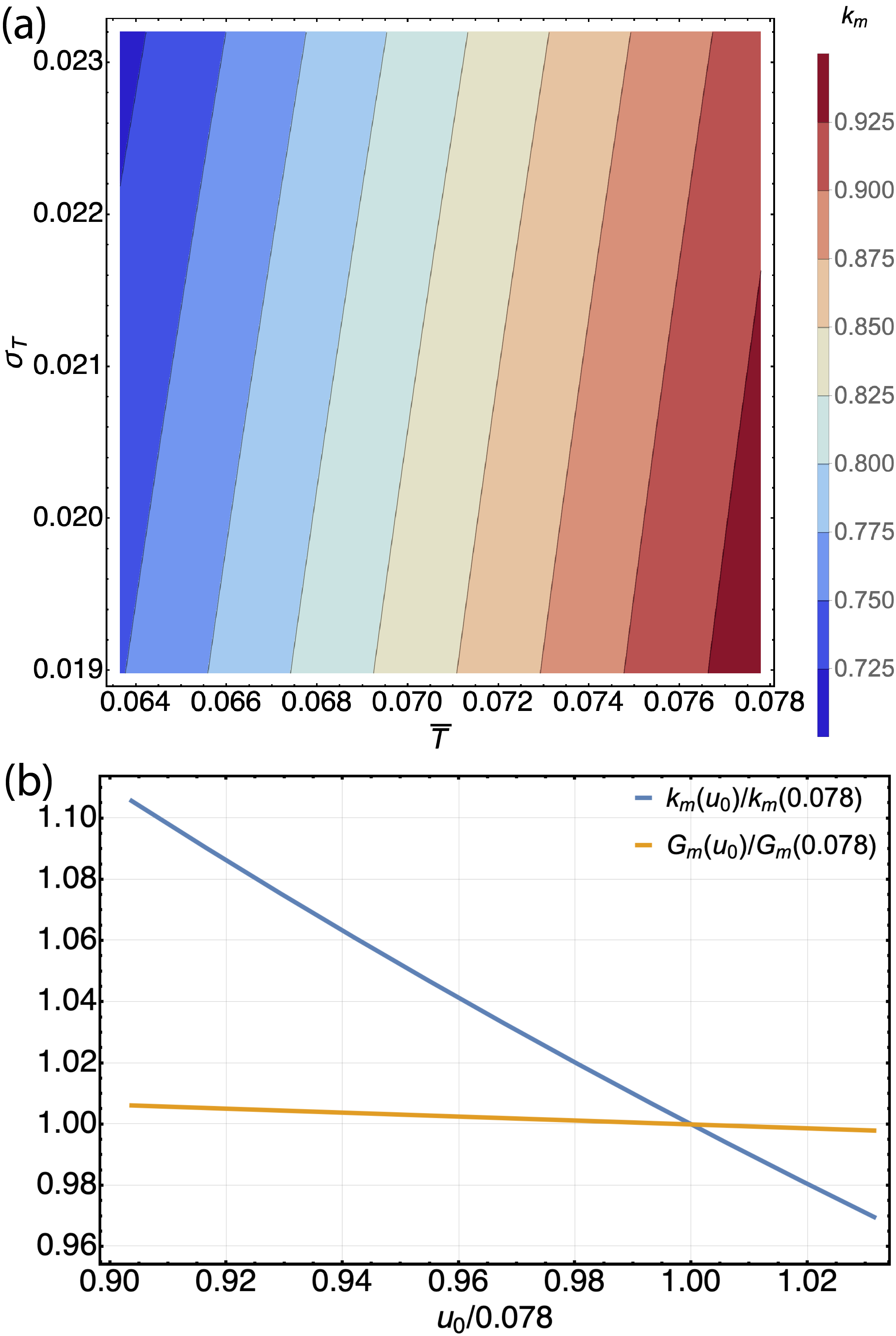}
\caption{(a) Density plot for $k_m$ as a function of $\bar{T}$ and $\sigma_T$ for a $P(T_\alpha)$ approximated by a Gamma distribution at a fixed $u_0=0.078$. The contour lines represent regions in \{$\bar{T}, \sigma_T$\} space with equal $k_m$, suggesting that a network with lower $\bar{T}$ but lower $\sigma_T$ can be stiffer than a network with higher $\bar{T}$ and $\sigma_T$. (b) For a fixed tension distribution ($\bar{T}=0.071$ and $\sigma_T = 0.021$), a $10\%$ reduction in $u_0$ results in about $10\%$ increase in $k_m$ and only a $0.7\%$ increase in $G_m$, suggesting that $k_m$ is sensitive to errors in $u_0$ estimation yet $G_m$ remains robust. The curves are normalized by the values of $k_m$ and $G_m$ at the actual $u_0$ measured from simulation, $u_0=0.078$.}
\label{fig:sensitivity}
\end{figure}

Next, we study the dependence of the EMT shear modulus on the model parameters. If we assume a gamma distribution for $P(T_\alpha)$ we find that $k_m$ increases with increasing $\bar{T}$ and decreasing $\sigma_T$ (Fig.~\ref{fig:sensitivity}a). An implication of this is that a network with low $\bar{T}$ but also low $\sigma_T$ can be stiffer than another network that has a larger $\bar{T}$ but also larger $\sigma_T$. Another important parameter is $u_0$, which may be challenging to accurately estimate in experiments. We thus check sensitivity of $G_m$ to errors in $u_0$ for a given $P(T_\alpha)$, and find that even though $k_m$ shows a high sensitivity, $G_m$ is robust against changes in $u_0$ (Fig.~\ref{fig:sensitivity}b). In fact, it would be sufficient for a given network to guess $u_0 = \bar{T}/k$. For instance, in the example given in Fig.~\ref{fig:sensitivity}b, choosing $u_0 = \bar{T}/k$ would only cause a $0.7\%$ error in $G_m$. The last parameter that appears in the EMT calculation is $L_0^*$ which independently affects $a^*$. We checked that if instead we used $L_{h.c.}=0.6204$ -- the edge length of honeycomb lattice if the area of its hexagons was equal to the average area of the polygons in the disordered network -- the resulting $G_m$ would change by less than $1\%$.  These tests suggest that $G_m$ does not have a strong explicit dependence on $u_0$ or $L_0^*$, and is determined mainly by $P(T_\alpha)$, which makes this approach more attractive for estimating stiffness of networks if the only data we have is $P(T_\alpha)$.

\begin{figure}
\centering
    \includegraphics[width=\linewidth]{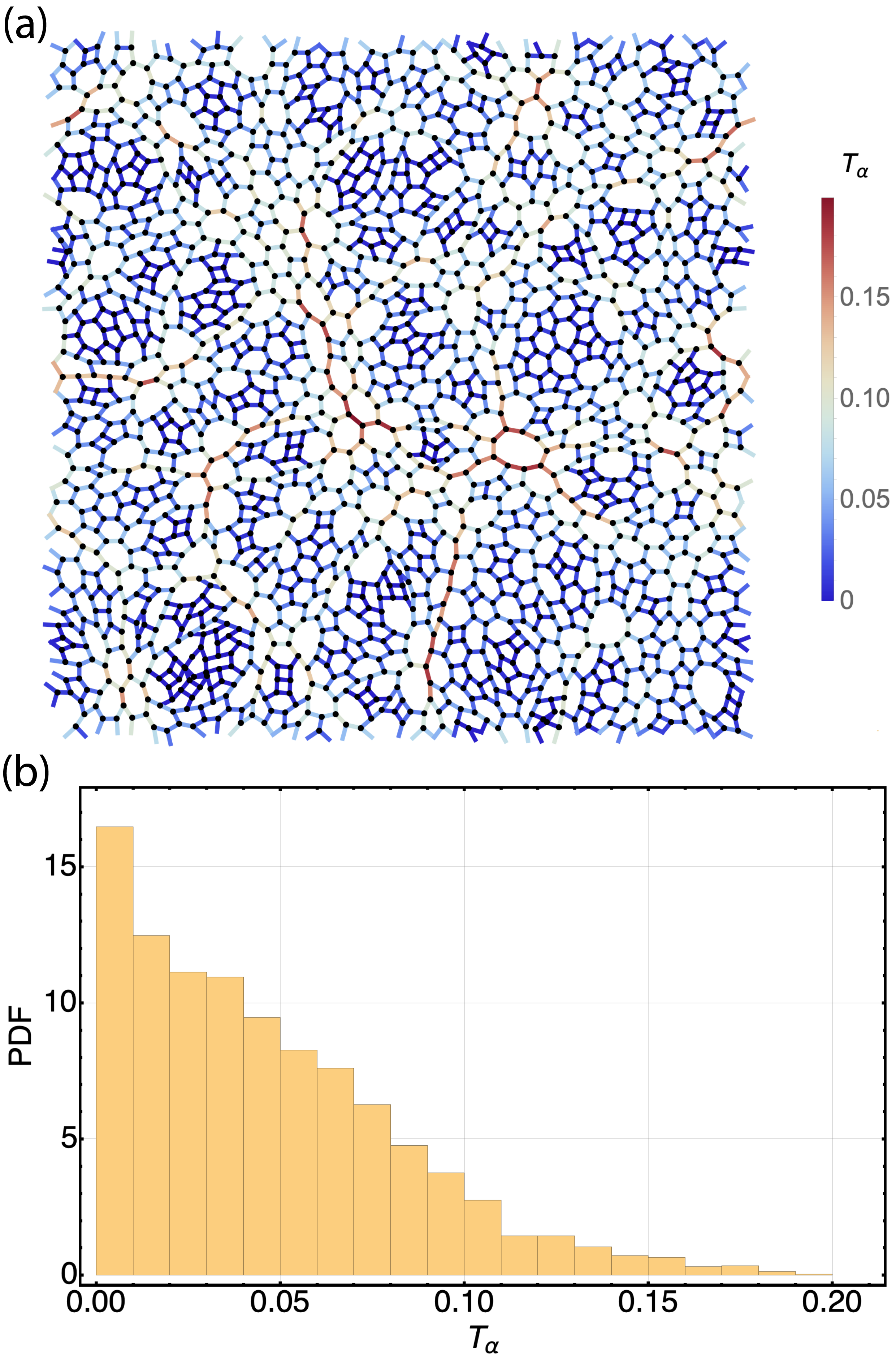}
\caption{(a) A non-Gamma rigid random regular network with $N=1936$ vertices and $L_0 = 0.600$ in a periodic box with $L_{box} = \sqrt{968}$ (for this network $L_0^*\approx0.680$). The colors show the tensions on the edges. (b) Probability distribution function (PDF) for tension distribution $P(T_\alpha)$ for the same network.}
\label{fig:spvFalse}
\end{figure}

So far, all generated random regular networks possess $P(T_\alpha)$ well-fit by Gamma distributions. By setting $T_{spv} = 0$, we can, however, generate networks with tension distributions that have very high spatial heterogeneities not captured by a Gamma distribution, e.g.\ rigid networks containing edges with zero or negative tensions. One could then ask how well the EMT predicts the shear modulus of such networks. Unfortunately, Eq.~\ref{eq:self_consistency} does not have a stable fixed point for negative tensions. To circumvent this issue, we remove edges with negative tensions from the aforementioned networks while keeping $z=3$ (see SM), such that the resulting network has edges with zero or positive tensions and $P(T_\alpha)$ does not resemble a Gamma distribution (Fig.~\ref{fig:spvFalse}). 

For networks with non-Gamma tension distributions, we consistently find that the $G_m$ is smaller than the actual $G$. For instance, for the representative network shown in Fig.~\ref{fig:spvFalse}, we estimate $G_m = 0.07$ while direct calculations show $G = 0.30$. This is in contrast to Gamma distributed networks for which the error of $G_m$ is typically less than $50\%$. We believe the reason for such high errors in EMT predictions for these non-Gamma networks is the high degree of spatial heterogeneity observed in them. In particular, from Fig.~\ref{fig:spvFalse} we can see that high tension edges, which are outnumbered by low or zero tension edges, are spatially organized in chains that support the structure the network, making the network more rigid than it would be if tensions from $P(T_\alpha)$ were distributed among edges completely randomly. Such spatial correlations are not captured in this EMT formalism and require higher order corrections, an interesting avenue for future work. Looking at the nematic order parameter $\bar{S}$ for the example in Fig.~\ref{fig:spvFalse}, we see that indeed this network has a higher $\langle S \rangle$ than network 1 and 2: $\bar{S}_3 = 0.515 \pm 0.015$ and thus has a higher degree of spatial correlations. See SM for more details.

\section{Discussion}
We have developed an Effective Medium Theory for density-independent rigidity transitions in random, regular networks. This approach empowers us to predict how, for example, the shear modulus increases with increasing strain once determining the distribution of network tensions. Interestingly, in many instances, this distribution is well-characterized by a Gamma distribution. Near the transition, the EMT predicts a linear dependence of the shear modulus on strain and is in agreement with our numerical calculations. This linear dependence is also consistent with an earlier mean field approach that did not invoke a self-consistency condition but did rely on the average and standard deviation of the tension distribution~\cite{Merkel2019}. The fact that the shear modulus increases with strain does not necessarily require an EMT approach, as it can be captured by simply estimating the modulus from that of a honeycomb with the same $u_0$. However, such a simple model carries no information about the tension distribution, which can nontrivially impact the modulus. It is then noteworthy that even though tension standard deviation also increases with strain, EMT is still able to correctly capture the increasing trend of shear modulus. However, our EMT does not accurately predict the magnitude of the proportionality coefficient between the shear modulus and the strain. While one does not necessarily anticipate a mean field theory being quantitatively accurate, we delve further into the source of the discrepancy by modifying the heterogeneity of the geometric disorder in the spring network to address the strengths and short-comings of the EMT in more detail.

Modifying the algorithm for building the spring network results in varying the geometric disorder and, hence, affects the distribution of network tensions. Qualitatively, one expects that a network with higher heterogeneity, or larger standard deviation, with a similar average tension leads to a smaller shear modulus. Indeed, the EMT captures that phenomenon.  Yet, the average tension also plays a role such that increasing the average, as well as the standard deviation, may lead to a stiffer network.  The EMT captures this phenomenon as well. On the other hand, when making quantitative comparisons between the EMT predictions and the numerical simulations of the shear modulus, we find that the discrepancy depends on the spatial correlations in the network. One source of spatial correlations is local nematic ordering, which we measure. For less spatially-correlated spring networks, i.e., less local nematic ordering, the EMT achieves better accuracy than for more spatially-correlated, or more local nematic ordering, spring networks, indicating that one must move beyond a mean field approach to more accurately prediction the mechanics. 

For those networks that are heterogeneous but lack spatially-correlated disorder, the EMT is a good starting point for making predictions. Moreover, we find that the obtained shear modulus is rather robust to errors in the internal strain, which makes the approach ideal for experiments where only the tensions need to be measured accurately. To move beyond a good starting point--to construct an analytical framework with fluctuations--becomes an even more pressing issue given the minimal, yet powerful, spring network framework in the form of two-body, or even multi-body, interactions. There exists a field theory for amorphous solids well in the rigid phase~\cite{DeGiuliPRL,DeGiuliPRE}. Can such an approach be extended to disordered solids near the rigidity transition? Alternatively, how does one go beyond an EMT by incorporating fluctuations in a perturbative, but controlled,  manner?  Indeed, we must find out the answers to these questions if we are to make more quantitatively accurate predictions about the mechanical properties of disordered living and nonliving matter.

\acknowledgments
OKD wishes to thank Amanda Parker for discussion. JMS and MLM acknowledge financial support from NSF-PoLS-2014192. We would also like to acknowledge support from the Simons Foundation No 454947 (MLM and OKD), and NSF-DMR-1951921 (MLM).
\\


\title{\Large\textbf{Supplementary materials}}
\section{Finite size effects in the rigidity of random regular networks}
\begin{figure*}
\centering
    \includegraphics[width=\linewidth]{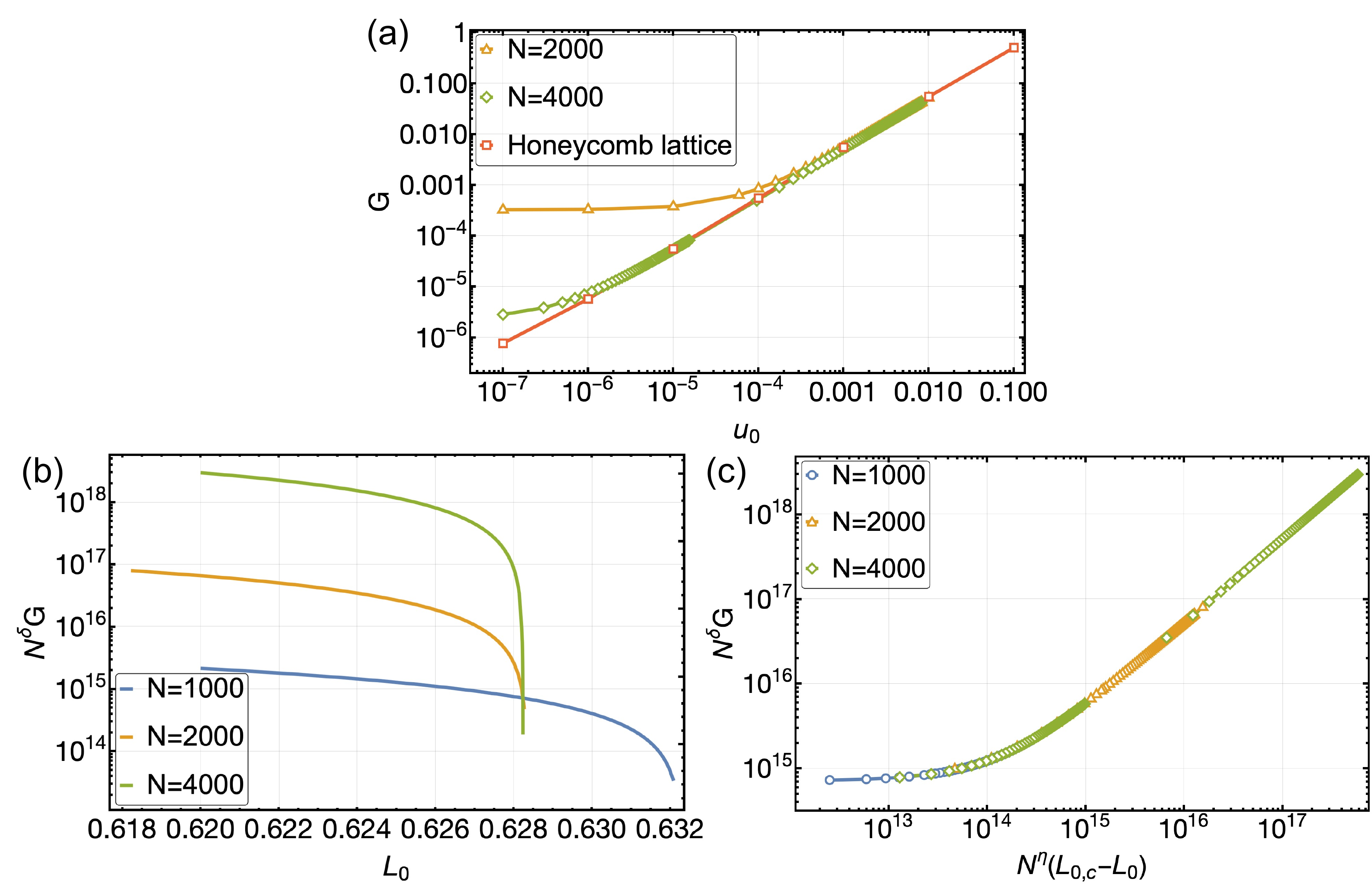}
\caption{Finite size scaling of random regular spring networks. (a) Shear modulus $G$ vs.\ $u_0 = L_0^* - L_0$ reaches a plateau near the rigidity transition point for finite-sized networks, causing a discontinuity at $L_0^* = L_0$ that appears to be reduced for larger networks. For comparison, the shear modulus of a honeycomb lattice is plotted, which shows a linear dependence on $u_0$. If the discontinuity in random regular networks is only due to finite size effects, it is expected to vanish in the limit of $N\to \infty$. (b) For $\delta \approx 5.51$, $N^\delta G$ vs.\ $L_0$ curves for different system sizes intersect at the same point numerically estimated to be $L_{0,c} \approx 0.628230$. (c) With $\eta \approx \delta \approx 5.51$ and $L_{0,c} \approx 0.628230$, $N^\delta G$ vs.\ $(L_{0,c}-L_0)N^\eta$ curves collapse into one as expected from finite size scaling.}  
\label{fig:finite-size-scaling}
\end{figure*}

In the main text, we alluded to a finite size effect observed in random regular networks at the rigidity transition. The effect manifests itself in a discontinuity in the shear modulus at the rigidity transition for finite networks (Fig.~\ref{fig:finite-size-scaling}a). Here we show by finite size scaling that this discontinuity is indeed due to finite size effects and is expected to vanish in the thermodynamic limit. We also numerically find the rigidity transition critical point $L_{0,c}$ in that limit.

We start by an ansatz that energy should scale like $E = E\left((L_{0,c} - L_0)b^{y_l},\gamma b^{y_\gamma},L_{box}^{-1}b\right)$ with scaling parameter $b$. Here, $L_{0,c}$ is the thermodynamic limit critical rest length, $\gamma$ is the shear parameter and $L_{box}=\sqrt{N/2}$ is the side length of the periodic box. We can find the scaling behavior of the shear modulus using $G$:
\begin{align}
    G &= \frac{1}{L_{box}^2}\frac{\mathrm{d}^2 E}{\mathrm{d}\gamma^2}|_{\gamma=0} \nonumber\\
    &= \frac{1}{L_{box}^2} b^{2y_\gamma}E_{\gamma \gamma}\left((L_{0,c} - L_0)b^{y_l},L_{box}^{-1}b\right).
\end{align}
Now, set $b=L_{box}$ to get
\begin{align}
    G &= L_{box}^{2y_\gamma-2}E_{\gamma \gamma}\left((L_{0,c} - L_0)L_{box}^{y_l},1\right)\nonumber\\
    &\sim N^{-\delta}g\left((L_{0,c} - L_0)N^\eta \right),
\end{align}
where $\delta = 1-y_\gamma$ and $\eta = y_l/2$.

If the discontinuity in $G$ is due to finite size effects, we should expect for the correct exponents $\delta$ and $\eta$:

1) $N^\delta G$ vs.\ $L_0$ curves for different system sizes should intersect at $L_{0,c}$.

2) $N^\delta G$ vs.\ $(L_{0,c}-L_0)N^\eta$ curves for different system sizes should collapse into one.

We generate three different system sizes: $N = 1000, 2000, 4000$ and find that $\delta \approx 5.51$ makes $N^\delta G$ vs.\ $L_0$ intersect at the same point, $L_{0,c} \approx 0.628230$ (Fig.~\ref{fig:finite-size-scaling}b). Then, using those $\delta$  and $L_{0,c}$ we see that for $\eta \approx \delta \approx 5.51$,  $N^\delta G$ vs.\ $(L_{0,c}-L_0)N^\eta$ curves collapse into one (Fig.~\ref{fig:finite-size-scaling}c). This demonstrates that the shear modulus discontinuity is due to finite system sizes and that in the thermodynamic limit, we should expect $G$ to increase linearly with strain as discussed in \cite{Merkel2019}.

\section{Numerical simulations of spring networks}
For random regular spring networks, we use cellGPU~\cite{Sussman2017} to initialize $N_{cell} = N/2$ cell centers randomly in a square, periodic box of volume $L_{box}^2 = N_{cell}$. A Voronoi tessellation is applied to identify $N_{cell}$ polygon cells with $2N_{cell}$ vertices with coordination number $z=3$. The final step in the initialization process involves moving the cell centers for a few time steps using a self-propelled Voronoi model \cite{Bi2016} to make cell areas more uniform (see below for more details). After the initialization process, the energy is minimized using the FIRE minimizer~\cite{Bitzek2006} with a force cutoff of $10^{-12}$. The time step for the simulations was dynamically decided by the minimizer, starting from $\mathrm{d}t=0.001$, but allowed to be increased up to $\mathrm{d}t_{max}=0.1$.

\section{Generating networks with different degrees of spatial heterogeneity}
After random placement of cell centers in the periodic box and performing a Voronoi tessellation, one can apply a self-propelled voronoi (SPV) dynamics to move cell centers for a few time steps so that the resulting network looks more uniform. The process is outlined in~\cite{Sussman2017} but we also give a brief overview here: In SPV dynamics, we assume the polygons are epithelial cells with an area elasticity and perimeter elasticity with energy $E = \sum_{\alpha = 1}^{N/2} [K_A (A_\alpha - A_0)^2 + K_P(P_\alpha - P_0)^2]$, where $K_A = 1.0$ and $K_P=1.0$ are area and perimeter elastic constants, $A_\alpha$ and $P_\alpha$ are the area and perimeter of cell $\alpha$, $A_0 = 1.0$ and $P_0=3.8$ are preferred area and preferred perimeter of cells. The dynamics of the $\alpha$'th cell center position $\mathbf{r}_\alpha$ is given by $\dot{\mathbf{r}}_\alpha = \mu \mathbf{F}_\alpha + v_0 \mathbf{n}_\alpha$ where $\mathbf{F}_\alpha = -\nabla_\alpha E$ is the force on cell center $\alpha$, $\mu = 1.0$ the mobility, $v_0=0.1$ the self propulsion speed and $\mathbf{n}_\alpha$ the direction of cell $\alpha$'s self propulsion which involves rotational noise $D_r = 1.0$ (see~\cite{Sussman2017} for more details). $dt = 0.1$ was chosen for the SPV dynamics. 

For networks 1 and 2, we ran the SPV dynamics for $T_{spv}=100$ and $T_{spv}=5$ respectively. Both networks had Gamma-like tension distributions. One could also have a network with $T_{spv}=0$. We observed that such networks have tension distributions that do not resemble Gamma distributions and even include edges with negative (compressive) tensions. To be able to perform EMT on these networks, we thus first removed negative tension edges with a few series of T1 and T2 transitions (ref.~\cite{Sussman2017}), confirming the final network still exhibited a non-Gamma tension distribution. 

\section{Measurement of nematic ordering}
To measure average nematic ordering in each network, we first define a $2D$ nematic tensor for an edge: $Q_{ij} = 2(n_i n_j - \delta_{ij}/2)$ where $n_i$ is the $i$'th component of the unit vector along the edge. To find the average nematic tensor $\bar{Q}_{ij}(\mathbf{x})$ in the vicinity of a point $\mathbf{x}$ in the box, we calculate $Q_{ij}$ for edges that fall inside a small circle of radius $r$ centered on $\mathbf{x}$. $\bar{Q}_{ij}(\mathbf{x})$ has two parameters: $\theta(\mathbf{x})$, the alignment direction, and $S(\mathbf{x})$, the degree of alignment (scalar nematic order parameter), thus, the eigenvalues of $\bar{Q}_{ij}(\mathbf{x})$ are $\pm S(\mathbf{x})$. 

\begin{figure*}
\centering
    \includegraphics[width=\linewidth]{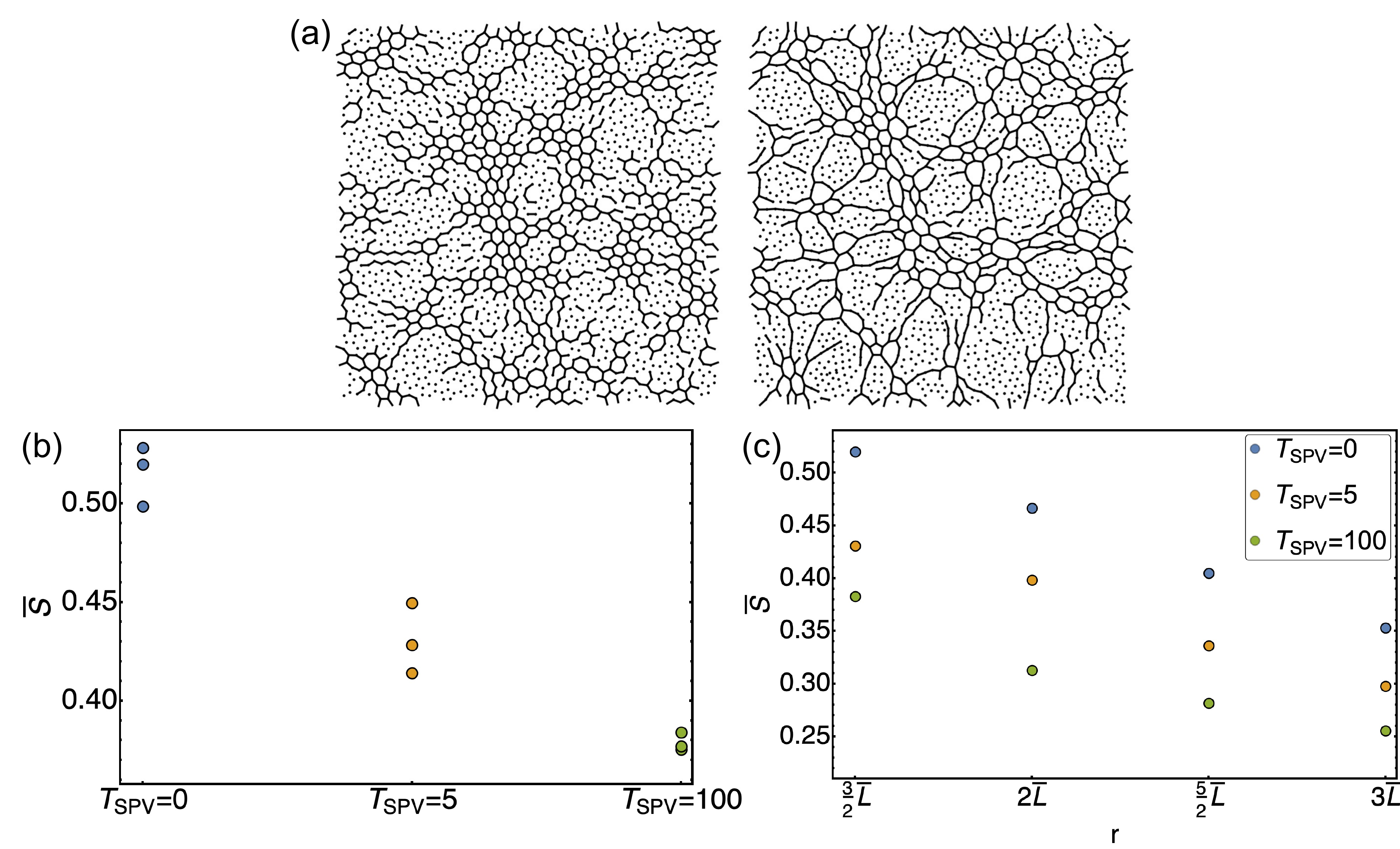}
\caption{Measurement of nematic ordering. (a) Visualization of the edges with tension higher than the average edge tension in a network with $T_{spv}=100$ (left) and $T_{spv}=0$ (right) shows significantly more chain-like structures in the $T_{spv}=0$ network. (b) Networks with higher spatial heterogeneity show higher nematic ordering. Here, $r=3\bar{L}/2$ where $\bar{L}$ is the average edge length in a network.  (c) $\bar{S}$ vs.\ $r$ for networks with different levels of spatial heterogeneity shows a reduction in $\bar{S}$. However, regardless of $r$, $T_{spv}=0$ consistently exhibits the largest $\bar{S}$ followed by $T_{spv}=5$ and then $T_{spv}=100$.}  
\label{fig:nematic-ordering}
\end{figure*}

To find an average nematic order parameter $\bar{S}$ for a network, we first remove all edges with tensions below the average tension in the network since high tension edges are the ones that hold the network together and show a correlated spatial structure. We then move the circle across the network (each time by $1.5 r$) to calculate $S(\mathbf{x})$ locally and then average over the whole network (only including circles that contained two or more edges). We picked $r$ such that roughly three edges are contained within the circle. We also observed that changing the radius does not change the overall result that networks with more spatial heterogeneity show higher nematic ordering (Fig.~\ref{fig:nematic-ordering}). 

\bibliographystyle{eplbib}
\bibliography{apssamp}





\end{document}